\documentclass[final,5p,times,twocolumn]{elsarticle}
\usepackage{graphicx}
\usepackage{amsmath}
\usepackage{geometry}
\usepackage{amssymb}

\journal{Optics Communications}

\begin{document}
	
\begin{frontmatter}

\title{Conditions for achieving invisibility of hyperbolic multilayered nanotubes}

\author{Carlos D\'{\i}az-Avi\~n\'o, Mahin Naserpour, and Carlos J. Zapata-Rodr\'{i}guez\corref{cor*}}
\cortext[cor*]{Corresponding author: carlos.zapata@uv.es}
\address{Department of Optics, University of Valencia, Dr. Moliner 50, Burjassot 46100, Spain}


\begin{abstract}
Highly-anisotropic plasmonic nanotubes exhibit a dramatic drop of the scattering cross section in the transition regime from hyperbolic to elliptic dispersion.
The characterization of a realistic multilayered metamaterial is typically carried out by means of an effective medium approach providing average components of the permittivity tensor and wave fields.
Here, the edge effects of the metal-dielectric stratified nanotube for different combinations were thoroughly analyzed.
We show how the boundary layers, which in principle remain fully irrelevant in the estimation of the effective permittivity of the nanotube, however play a critical role in the resonant scattering spectra and the near field patterns.
A remarkable enhancement of the scattered wave field is unexpectedly experienced at the frequencies of interest when a dielectric layer is chosen to be in contact with the cavity core.
\end{abstract}


\end{frontmatter}

\section{Introduction}

Metamaterials with unprecedented optical properties have been recently proposed by several groups for applications in cloaking and invisibility.
A vast majority of cloaking nanostructures are engineered in the basis of transformation optics principles \cite{Leonhardt06,Pendry06,Cai07} enabling analytical expressions of the electromagnetic fields and metamaterial properties \cite{Forouzeshfard15,Yu15,Rajput16}.
An original concept suggested by Alu and Engheta has also attracted great attention which relies on the use of metamaterial (or metal) coatings to severely drop the scattering efficiency of an object by means of a nonresonant scattering-cancellation approach \cite{Alu05}.
The strategy of such scattering cancellation takes advantage of the local negative polarizability of metamaterials, and its experimental realization was first demonstrated at microwave frequencies by using an array of metallic fins which are embedded in a high-permittivity environment to generate a metamaterial cloaking shell \cite{Edwards09}.
 
Novel extensions of the previous concepts have more recently made the scene, which are based on the use of double-shell and multilayered plasmonic coatings \cite{Tricarico09,Zhou16}.
For instance, using plasmonic shells with an epsilon-near-zero material enables to reduce substantially the scattering losses and simultaneously providing the shielding of the cloaked region \cite{Alu08,Filonov12}.
As an alternative, a nanotube consisting of a periodic distribution of metal and dielectric alternating layers, where the stratified metamaterial was described as a radial-anisotropic hyperbolic medium, has recently demonstrated narrow-band ultra-low scattering \cite{Kim15}.
The invisibility spectral band occurs when one of the components of the effective permittivity tensor is near zero.
The effective medium theory was adopted to efficiently reproduce the results provided by the analytical Lorenz-Mie scattering method \cite{Bussey75,Bohren98}.

Here, we study in detail the edge effects of the stratified hyperbolic nanotube employed in Ref.~\cite{Kim15} for different combinations, that is, when the layer in contact with the environment medium is either the metal or the nonconducting material.
We demonstrated that boundary effects play a relevant role in the resulting scattering efficiency of the nanotube, potentially clearing away the characteristic invisibility of the nanotube and boosting additional plasmonic resonances in the visible.
Explicitly, when silver is set as the constituent material of the outermost layer, unexpectedly, a significant enhancement of the scattered signal is observed.
As a consequence, the effective medium theory enabling a simplified model for the optical characterization of the nanoparticle may apparently lead to fallacious estimations even for metamaterials composed of subwavelength layers with a few-nanometers width.

\section{The hyperbolic multilayered nanotube}

\begin{figure}[tb]
\centering
\includegraphics[width=.7\linewidth]{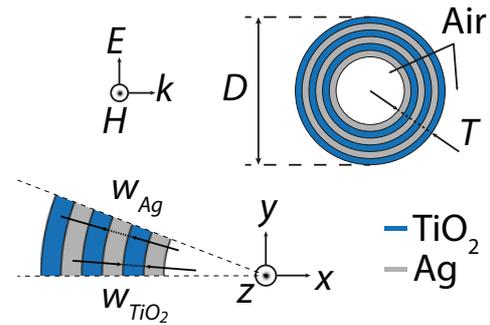}
\caption{Illustration of the multilayered nanotube.}
\label{fig01}
\end{figure}

\begin{figure*}[tb]
	\centering
	\includegraphics[width=.9\linewidth]{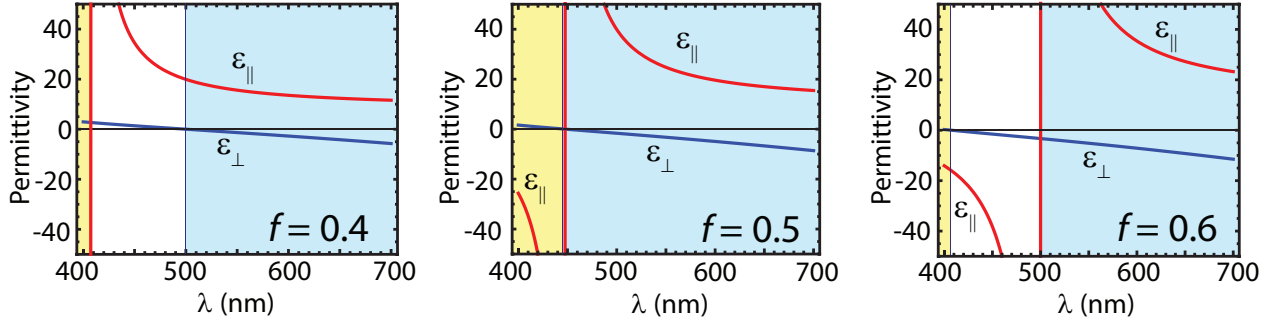}
	\caption{Real part of the components $\epsilon_\parallel$ (red line) and $\epsilon_\perp$ (blue line) of the permittivity tensor in the visible regime for different metal filling factors.
		The shaded regions denote spectral bands where the metamaterial exhibits a hyperbolic dispersion of the Type I (shaded in yellow) and the Type II (shaded in blue).}
	\label{fig02}
\end{figure*}

Let us consider a cylindrical shell formed by a multilayered nanostructure as illustrated in Fig.~\ref{fig01}.
Following Ref.~\cite{Kim15}, silver and titanium dioxide was used for the stratified medium.
A silver nanofilm of width $w_\mathrm{Ag}$ is set by the side of a TiO$_2$ layer of width $w_\mathrm{TiO_2}$, thus forming the unit cell of periodic distribution along the radial coordinate.
In our numerical simulations we considered a nanostructured tube of outermost diameter $D$ and composed of six layers, giving a total shell thickness of $T = 3 (w_\mathrm{Ag} + w_\mathrm{TiO_2})$.
The relative permittivities of silver and TiO$_2$ within the visible range of frequencies may be approximated by \cite{Kim15}
\begin{equation}
 \epsilon_\mathrm{Ag}(\lambda) = 3.691 - \frac{9.152^2}{(1.24/\lambda)^2 + i 0.021 (1.24/\lambda)} ,
 \label{eq01}
\end{equation}
and
\begin{equation}
 \epsilon_\mathrm{TiO_2}(\lambda) = 5.193 + \frac{0.244}{\lambda^2 - 0.0803} ,
 \label{eq02}
\end{equation}
respectively.
In the previous equations, the working wavelength $\lambda$ is set in micrometers.
In this study we examined hollow cylinders immersed in air.

Within a long-wavelength approach, the multilayered metamaterial may be considered as a uniaxial medium whose optic axis is set along the direction of periodicity, that is the radial axis in our particular case.
The effective anisotropic medium is then optically characterized by a local permittivity tensor $\underline{\epsilon}$ of component \cite{Yeh88}
\begin{equation}
 \epsilon_\parallel = \frac{\epsilon_\mathrm{Ag}\epsilon_\mathrm{TiO_2}}{f \epsilon_\mathrm{TiO_2} + (1 - f) \epsilon_\mathrm{Ag}} ,
 \label{eq03}
\end{equation}
along the optic axis, and 
\begin{equation}
 \epsilon_\perp = f \epsilon_\mathrm{Ag} + (1 - f) \epsilon_\mathrm{TiO_2} ,
 \label{eq04}
\end{equation}
in the perpendicular direction.
In the previous equations, the metal filling factor 
\begin{equation}
 f = \frac{w_\mathrm{Ag}}{w_\mathrm{Ag} + w_\mathrm{TiO_2}} ,
 \label{eq05}
\end{equation}
represents the unique geometrical parameter determining the effective permittivities of the metamaterial, regardless the internal one-dimensional distribution of the materials involved in the unit cell.

In Fig.~\ref{fig02} we represent the real values of $\epsilon_\parallel$ and $\epsilon_\perp$ within the visible range of frequencies, for different values of the metal filling factor.
We observe that the real part of $\epsilon_\perp$ is near zero around a filling factor $f_z = \epsilon_\mathrm{TiO_2} / (\epsilon_\mathrm{TiO_2} - \mathrm{Re} (\epsilon_\mathrm{Ag}))$, whereas the real part of $\epsilon_\parallel$ is near a pole around a filling factor $f_p = \mathrm{Re} (\epsilon_\mathrm{Ag}) / (\mathrm{Re} (\epsilon_\mathrm{Ag}) - \epsilon_\mathrm{TiO_2})$.
The hyperbolic regime is determined by the condition $\mathrm{Re} (\epsilon_\parallel) \mathrm{Re} (\epsilon_\perp) < 0$, indicated as shaded regions in Fig.~\ref{fig02}.
The choice $\mathrm{Re} (\epsilon_\perp) > 0$ corresponds to the so-called Type I hyperbolic metamaterials, whereas the choice $\mathrm{Re} (\epsilon_\perp) < 0$ denotes a Type II hyperbolic medium \cite{Ferrari15}. 
Note that when $\mathrm{Re} (\epsilon_\mathrm{Ag}) = - \epsilon_\mathrm{TiO_2}$ occurring at a wavelength $\lambda = 448$~nm, both $f_z$ and $f_p$ coincide at a value $0.5$, denoting that the silver layers and the TiO$_2$ layers have the same width.
In this case, the hyperbolic regime spans the whole spectrum.
Let us point out that the extraordinary dispersion features of hyperbolic and epsilon-near-zero metamaterials have inspired us in a plethora of applications such as subwavelength imaging \cite{Zapata11e,Zapata12a}, surface-wave engineering \cite{Miret10,Zapata13b}, and double refraction \cite{Zapata14b,Diaz16}, to mention a few.

\section{Results and discussion}

\begin{figure}[tb]
	\centering
	\includegraphics[width=.8\linewidth]{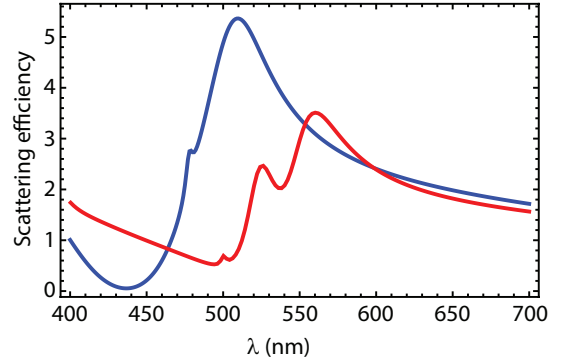}
	\caption{
		Scattering efficiency ($Q_s$) of a hollow metamaterial cylinder composed of 6 layers, where the nanotube thickness is $T = 60$~nm, its diameter is $D = 220$~nm, and the silver filling factor is $f = 0.5$.
		The inmost layer is made of either silver (blue line) or titanium dioxide (red line).
		The calculations are performed in the visible spectral range for TE$^z$-polarized incident light.
	}
	\label{fig03}
\end{figure}

\begin{figure*}[tb]
	\centering
	\includegraphics[width=.9\linewidth]{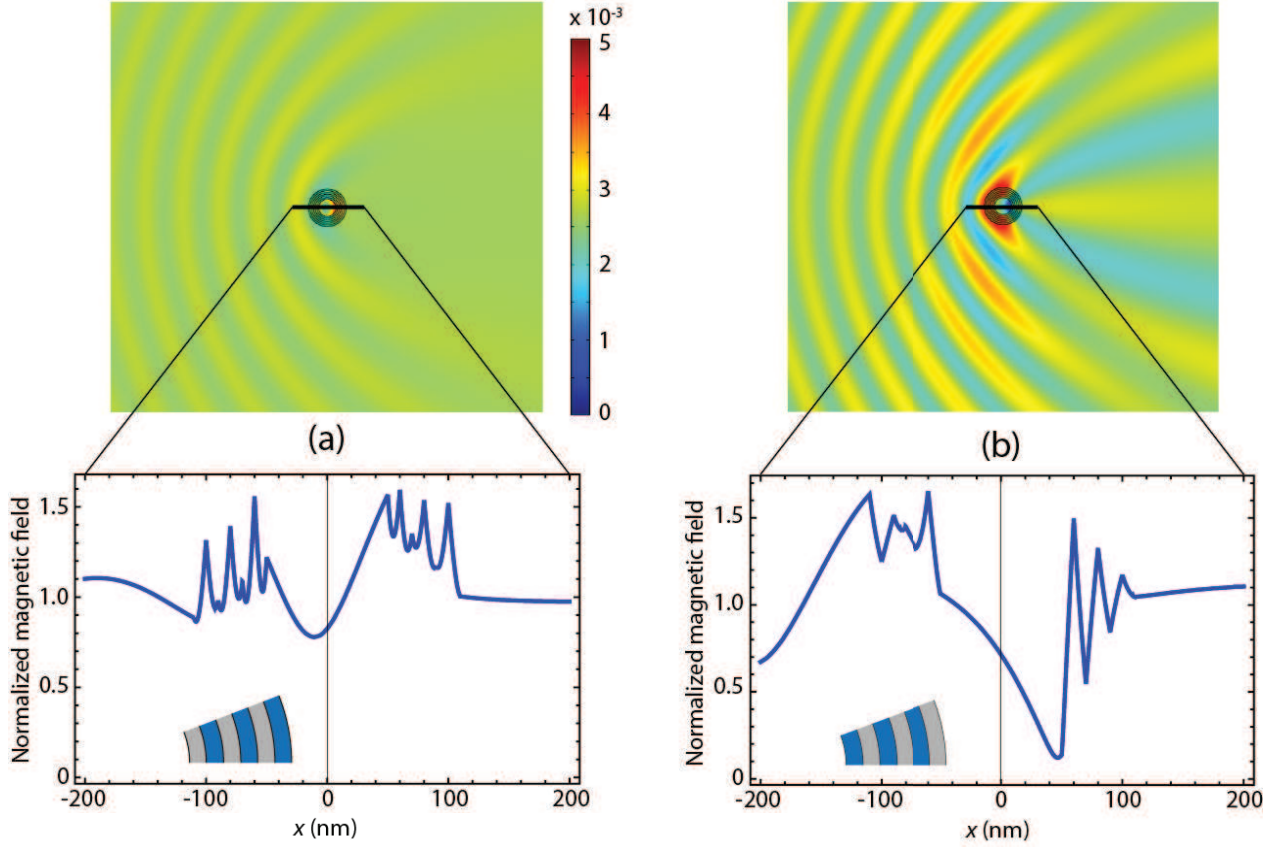}
	\caption{Amplitude distribution of the normalized magnetic field $|\mathbf{H}|/|H_0|$ at the invisibility wavelength $\lambda = 437$~nm and different arrangements of the metamaterial; 
		The inmost layer is made of (a) Ag, and (b) TiO$_2$.}
	\label{fig04}
\end{figure*}

To estimate analytically the scattering efficiency of the multilayered Ag-TiO$_2$ nanotube we followed the Lorenz-Mie scattering method \cite{Bussey75}.
Note that the axis of the cylindrical cavity is oriented along the $z$ axis.
First we assumed that the nanotube is illuminated by a TE$^z$-polarized plane wave propagating along the $x$ axis, as illustrated in Fig.~\ref{fig01}.
The magnetic field of the incident plane wave may be set as 
\begin{equation}
 \mathbf{H}^i = \hat{z} H_0 \exp \left( i k x \right) ,
\end{equation}
where $H_0$ is a constant amplitude and $k = 2 \pi / \lambda$ is the wavenumber in air.
In this case, the scattered magnetic field may be set as
\begin{equation}
 \mathbf{H}^s = \hat{z} H_0 \sum_{n = - \infty}^{+\infty} a_n i^n H_n^{(1)} \left(k r \right) \exp \left( i n \phi \right) ,
 \label{eq06}
\end{equation}
where $r$ and $\phi$ are the radial and azimuthal cylindrical coordinates, respectively, and $H_n^{(1)}$ is the Hankel function of the first kind and order $n$.
By using the Lorenz-Mie theory \cite{Bohren98}, it is possible to evaluate analytically the scattering coefficients $a_n$, which provide the estimation of the scattering efficiency as
\begin{equation}
 Q_s = \frac{4}{k D} \sum_{n = - \infty}^{+\infty} |a_n|^2 .
 \label{eq07}
\end{equation}
Plasmonic resonances are determined by the poles of the coefficients $a_n$.
On the other hand, the invisibility condition is established provided that the scattering coefficients $a_n$ arrive simultaneously to a value near zero.

\begin{figure}[tb]
	\centering
	\includegraphics[width=.8\linewidth]{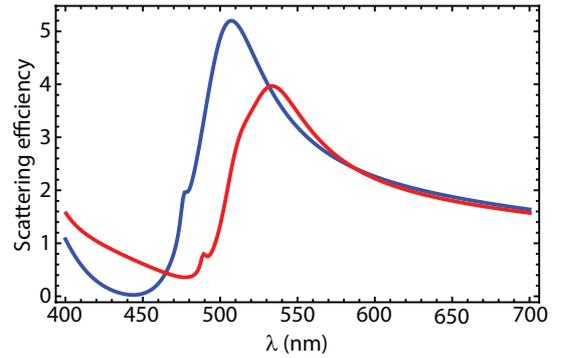}
	\caption{Scattering efficiency of a hollow metamaterial cylinder composed of 12 layers, where $T = 60$~nm, $D = 220$~nm, and $f = 0.5$.
		The interior layer is made of either titanium dioxide (red line) or silver (blue line).}
	\label{fig05}
\end{figure}

\begin{figure*}[h!]
	\centering
	\includegraphics[width=.9\linewidth]{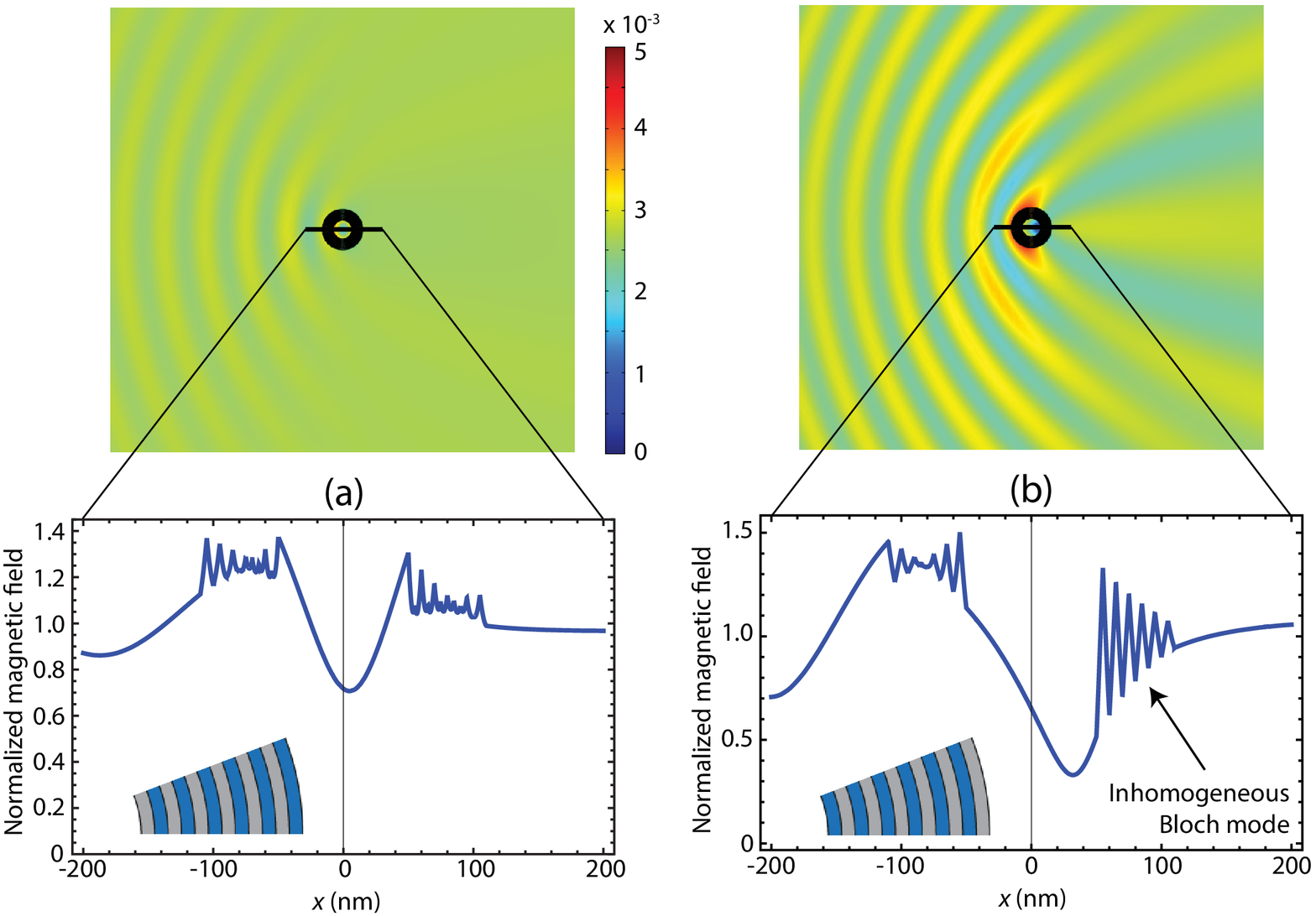}
	\caption{Amplitude distribution of the normalized magnetic field $|\mathbf{H}|/|H_0|$ at the invisibility wavelength $\lambda = 444$~nm and different arrangements of the metamaterial, where $T = 60$~nm, $D = 220$~nm, and $f = 0.5$.
		The interior layer is made of (a) silver, and (b) titanium dioxide.}
	\label{fig06}
\end{figure*}

In Fig.~\ref{fig03} we show the scattering efficiency of our Ag-TiO$_2$ nanotube calculated with Eq.~(\ref{eq07}), provided that the hyperbolic metamaterial has a metal filling factor $f = 0.5$ and that the hollow particle is illuminated by a TE$^z$ polarized plane wave.
In the numerical simulations we set a layer width $w_\mathrm{Ag}$ and $w_\mathrm{TiO_2}$ of 10~nm and a cylinder diameter of $D = 220$~nm.
Equation~(\ref{eq07}) is accurately estimated by using a finite number of scattering coefficients satisfying $|n| < 5$.
As illustrated in Fig.~\ref{fig03}, the calculated $Q_s$ exhibits significant discrepancies when the inmost layer is either Ag or TiO$_2$, despite $\epsilon_\parallel$ and $\epsilon_\perp$ remain the same in both cases.
When the interior layer is made of silver, the scattering efficiency of the nanostructure reaches a minimum of $0.053$ at $\lambda = 437$~nm, as reported in Ref.~\cite{Kim15}, where $\mathrm{Re} (\epsilon_\perp) = 0.34$ and $\mathrm{Re} (\epsilon_\parallel) = -143$.
In addition, a high peak of $Q_s = 5.36$ is found at $\lambda = 510$~nm, where $\mathrm{Re} (\epsilon_\perp) = -1.96$ and $\mathrm{Re} (\epsilon_\parallel) = 35$.
However, the number of resonant peaks increases if titanium dioxide is set in the inmost layer.
In this case, the highest peak is located at a wavelength of 560~nm where the scattering efficiency yields $3.50$, and a secondary strong peak is found at $\lambda = 525$~nm ($Q_s = 2.46$).
In addition, its minimum in scattering efficiency is shifted at $\lambda = 491$~nm where importantly it reaches a value of $0.53$.
We point out that such minimum in scattering efficiency, which is far of being associated with invisibility, is one order of magnitude higher than that estimated previously for an Ag inmost coating.
Finally, a dissimilar behavior in scattering is also manifested for TM$^z$-polarized incident light (not shown in Fig.~\ref{fig03}). 

In order to provide a physical insight of such a behavior, we analytically calculated the electromagnetic fields inside the nanotube and in the air core and in the environment medium.
In Fig.~\ref{fig04} we represent the modulus of the magnetic field, $|\mathbf{H}|$, at the invisibility wavelength $\lambda = 437$~nm for different geometric configurations, calculated by means of the Lorenz-Mie scattering method, provided that the incident wave field is TE$^z$ polarized.
When silver is set in the inmost layer of the hyperbolic metamaterial, we observe a moderate enhancement of the field at three different Ag-TiO$_2$ interfaces, as shown in Fig.~\ref{fig04}(a), which corresponds to a collective excitation of surface plasmon polaritons (SPPs).
It is clear that the field exhibits strong irregularities derived from nonlocal effects in the multilayered metal-dielectric nanostructure \cite{Elser07}.
Though such phenomenon severely puts into question the validity of the effective medium approach followed in Ref.~\cite{Kim15}, the deviations from an average field inside the hyperbolic nanotube are moderate in this case.
Analyzing the case that titanium dioxide is placed in contact with the interior medium, however, we observe even abrupter variations of the field especially at the rear part of the nanotube (in the semi-space $x > 0$), as shown in Fig.~\ref{fig04}(b).
Furthermore, now the field distribution is clearly asymmetric with respect to the origin of coordinates, leaving a strong backscattered signal.

\begin{figure*}[tb]
	\centering
	\includegraphics[width=.9\linewidth]{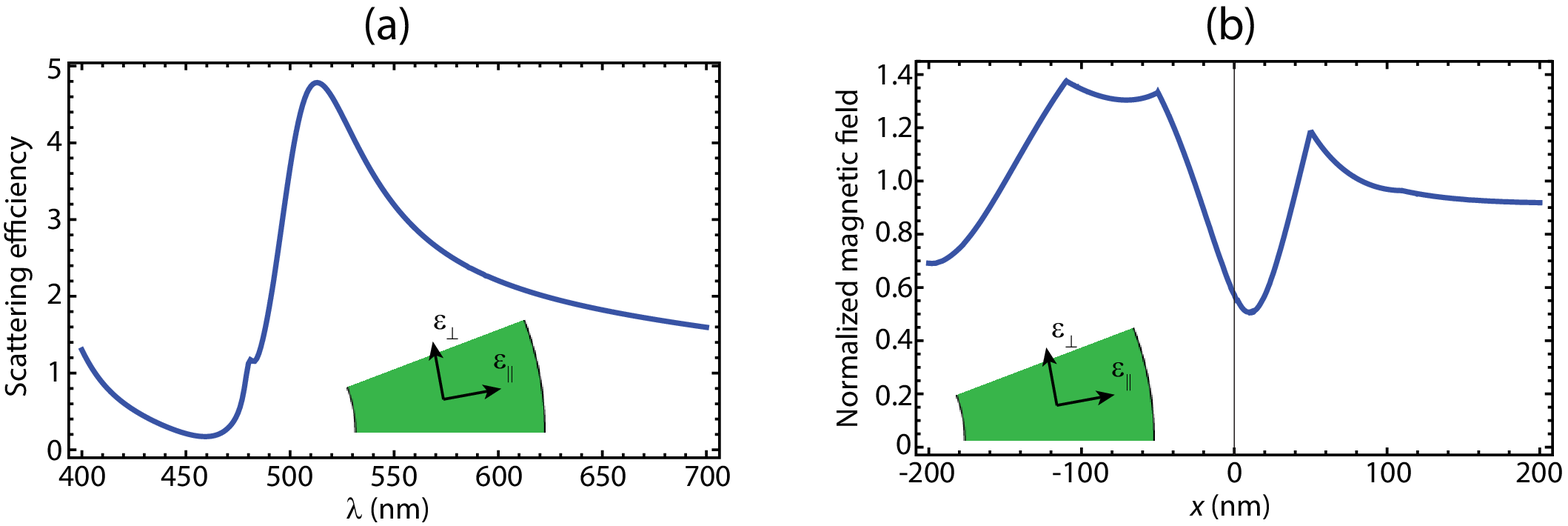}
	\caption{
		(a) Scattering efficiency of a hollow cylinder of thickness $T = 60$~nm and total diameter $D = 220$~nm, composed of a hyperbolic metamaterial with components of its permittivity tensor is given by Eqs.~(\ref{eq03}) and (\ref{eq04}), where $f = 0.5$.
		(b) Normalized magnetic field $|\mathbf{H}|/|H_0|$ along the $x$-axis at the invisibility wavelength $\lambda = 459$~nm.}
	\label{fig07}
\end{figure*}

Finally, we observe from Fig.~\ref{fig04} that the field amplitude at the center of the nanotube and in its neighborhood is, in average, near the field remaining in the environment medium rather than zero.
This fact evidences that the shielding effect that commonly is recognized in epsilon-near-zero core-shell nanocylinders \cite{Filonov12} cannot be reproduced in our hyperbolic metamaterial nanotube when the light impinges under TE$^z$ polarization.

The following analyzes the effects of increasing the number of layers in our nanotube maintaining the thickness $T = 60$~nm and the metal filling factor $f = 0.5$ of the shell, implying that the layer width $w_\mathrm{Ag}$ and $w_\mathrm{TiO_2}$ of the metamaterial is reduced.
Considering 12 alternating layers of silver and titanium dioxide, the layer width (and the period of the multilayered nanostructure) is reduced to one half with respect to the cases analyzed above and illustrated in Figs.~\ref{fig03} and \ref{fig04}.
The scattering efficiency of such additional nanotube is shown in Fig.~\ref{fig05}, considering the examples where the inmost layer is either Ag or TiO$_2$; the efficiency $Q_s$ is again evaluated by means of the Lorenz-Mie scattering method.
When the layer in contact with the core is silver, we find a minimum of $Q_s$ equal to $0.026$ at a wavelength $\lambda = 444$~nm, representing an exiguous variation in comparison with the hyperbolic nanotube of 6 layers.
In addition, the maximum of scattering efficiency reaches a value of $5.20$ at $\lambda = 507$~nm. 
If now we consider a nanotube with the interior layer made of TiO$_2$, its scattering efficiency at the invisibility wavelength $\lambda = 444$~nm reaches a value of $0.671$, representing a factor of 25 with respect to the case previously analyzed.
In fact, the minimum of the scattering efficiency is $0.345$, which is found at $\lambda = 477$~nm.
The overall shape of $Q_s$, however, becomes relatively similar in both cases.

The profile of the magnetic field at the new invisibility wavelength $\lambda = 444$~nm again exhibits abrupt changes at the metal-dielectric interfaces, as shown in Fig.~\ref{fig06}(a) for a nanotube with inmost layer made of Ag, but the envelope enabling the estimation of an average value presents a slow variation inside the metamaterial.
Moreover, a minimum of intensity is found near the origin of coordinates, however such intensity is far of zero disabling the gain of a shielding effect.
In Fig.~\ref{fig06}(b) we represent the magnetic field of the nanotube at the invisibility wavelength of $\lambda = 444$~nm, considering a nanotube of interior layer made of TiO$_2$.
The resonant excitation of collective SPPs at the rear part of the nanocylinder, developing a chief (inhomogeneous) Bloch wave in the multilayered metamaterial \cite{Torrent09}, seems to be responsible of the significant increase in the scattering efficiency and subsequently the predominant visibility of the nanotube.
Now this is clear that the appearance of a predominant radially-evanescent Bloch wave also occurs for a 6-layer nanostructure, as shown in Fig.~\ref{fig04}(b).

Obviously, by including a higher number of layers and, at the same time, maintaining the thickness $T$ of the nanotube and the filling factor $f$ of the metamaterial, both the scattering efficiency spectrum and the field pattern will continuously approach for the Ag and TiO$_2$ ending layers, leading to the response of a purely hyperbolic nanotube \cite{Kim15}.
Figure~\ref{fig07}(a) shows the scattering efficiency $Q_s$ of a hollow nanocylinder of thickness $T = 60$~nm and total diameter $D = 220$~nm, which is composed of a hyperbolic metamaterial with components of its permittivity tensor as given by Eqs.~(\ref{eq03}) and (\ref{eq04}), considering that the filling factor $f = 0.5$.
The calculations were carried out by using the Lorenz-Mie scattering method described in Refs.~\cite{Chen12b,Chen13,Chen15}.
The minimum efficiency ($Q_s = 0.172$) is found at $\lambda = 459$~nm establishing the invisibility wavelength, where $\mathrm{Re} (\epsilon_\perp) = -0.36$ and $\mathrm{Re} (\epsilon_\parallel) = 151$.
The main efficiency peak reaching $4.76$ is localized at $\lambda = 513$~nm.
The field distribution in the nanotube is also plotted in Fig.~\ref{fig07}(b) at the invisibility wavelength $\lambda = 459$~nm.
We observe the formation of surface waves in the two air-hyperbolic metamaterial interfaces. 
In average, this field pattern is closer to the field distribution shown in Fig.~\ref{fig06}(a) for an Ag inmost layer than that exhibited for a nanostructured tube of Ag ending coating depicted in Fig.~\ref{fig06}(b).
At the rear part of the nanotube, the field oscillations inside the multilayered metamaterial becomes stronger near the core; however, a surface wave at the metamaterial-core interface is only produced when a silver layer remains in contact with the core, as shown in Fig.~\ref{fig06}(a) but not in Fig.~\ref{fig06}(b).

\section{Conclusions}

In summary, we investigated numerically the scattering efficiency, with emphasis in the invisibility regime, of multilayered Ag-TiO$_2$ nanotubes with configurations of a different ending layer.
We demonstrated that when the nonconducting layer remains in contact with the core of the hollow cylinder, the characteristic invisibility of the hyperbolic nanotube is cleared away, even considering elementary layers of a few nanometers.
In this case there is an additional boosting of plasmonic-Bloch resonances markedly observed on the wave fields localized at the rear part of the multilayered nanocylinder.
Furthermore, the effective medium theory enabling a simplified model for the optical characterization of the nanoparticle will lead to fallacious estimations.
Such effect becomes weaker for metamaterials composed of subwavelength layers with less-than-5-nanometers width.

\section*{Acknowledgments}

This work was supported by the Spanish Ministry of Economy and Competitiveness (MINECO) (TEC2014-53727-C2-1-R).

\bibliographystyle{elsarticle-num} 

\end{document}